\documentclass{aastex631}
\usepackage{float}

\shorttitle{Spectroscopic distances, masses and ages}
\shortauthors{Stone-Martinez.}

\graphicspath{{./}{figures/}}
\usepackage{url}
\begin{document}

% \title{Parameter based stellar distances and masses using simple neural nets}
\title{Spectroscopic distance, mass, and age estimations for APOGEE DR17}

\author[0000-0003-4761-9305]{Alexander Stone-Martinez }
\affiliation{Department of Astronomy, New Mexico State University, P.O.Box 30001, MSC 4500, Las Cruces, NM, 88033, USA}

\author[0000-0002-9771-9622]{Jon A. Holtzman}
\affiliation{Department of Astronomy, New Mexico State University, P.O.Box 30001, MSC 4500, Las Cruces, NM, 88033, USA}

\author[0000-0003-2025-3585]{Julie Imig} 

\affiliation{Space Telescope Science Institute, 3700 San Martin Drive, Baltimore, MD 21218, USA}

\affiliation{Department of Astronomy, New Mexico State University, P.O.Box 30001, MSC 4500, Las Cruces, NM, 88033, USA}

\author[0000-0003-4752-4365]{Christian Nitschelm}
\affiliation{Centro de Astronomía (CITEVA), Universidad de Antofagasta, Avenida Angamos 601, Antofagasta 1270300, Chile}

\author[0000-0002-3481-9052]{Keivan G. Stassun}
\affiliation{Department of Physics and Astronomy, Vanderbilt University, VU Station 1807, Nashville, TN 37235, USA}

\author[0000-0002-8725-1069]{Joel R. Brownstein}
\affiliation{Department of Physics and Astronomy, University of Utah, 115 S. 1400 E., Salt Lake City, UT 84112, USA}

\begin{abstract}
We derive distances and masses of stars from the Sloan Digital Sky Survey (SDSS) Apache Point Observatory Galactic Evolution Experiment (APOGEE) Data Release 17 (DR17) using simple neural networks. Training data for distances comes from Gaia EDR3, supplemented by literature distances for star clusters. For masses, the network is trained using asteroseismic masses for evolved stars and isochrone masses for main sequence stars. The models are trained on effective temperature, surface gravity, metallicity and carbon and nitrogen abundances. We found that our distance predictions have median fractional errors that range from $\approx 20\%$ at low log g and $\approx 10\%$ at higher log g with a standard deviation of $\approx 11\%$. The mass predictions have a standard deviation of $\pm 12\%$. Using the masses, we derive ages for evolved stars based on the correspondence between mass and age for giant stars given by isochrones. The results are compiled into a Value Added Catalog (VAC) called DistMass that contains distances and masses for 733901 independent spectra, plus ages for 396548 evolved stars.
\end{abstract}

\section{Introduction} 
\label{sec:intro}
The Milky Way offers unique opportunities for studying galaxy formation and evolution. We can measure many of the properties of the nearby stars, such as the three dimensional position and velocity (through astrometry and spectroscopy), stellar parameters and stellar elemental abundances (through high-resolution spectroscopy), and stellar masses through asteroseimology and binary stars. However, these properties become more challenging to measure at larger distances from the Sun and for dimmer stars, so methods of extending the range of distances and luminosity over which we can reliably measure them is an important step to furthering our understanding of Galactic formation and evolution. 

Recently, large scale surveys such as APOGEE (\citealt{2017AJ....154...94M,2017AJ....154...28B}), GALAH \citep{Buder_2021}, LAMOST \citep{Wang_2021}, and Gaia-ESO \citep{Binks_2022} have increased the quantity of high-quality stellar parameters, Gaia \citep{2016A&A...595A...1G} has provided precise astrometric data in the form of parallaxes and proper motions for large swaths of the Galaxy, and light curve data from the Kepler space telescope has allowed for asteroseismic measurements of thousands of stars (e.g. \citealt{2018}). With asteroseismology it is possible to derive the evolutionary state, mean density, mass, and radius of stars. However, accurate parallaxes are more difficult to derive for stars at large distances, for example, at 5 kpc the median error in Gaia data is 15\% and at 8 kpc it grows to 34\%. Asteroseismic measurements are limited to the few distinct fields that were observed over long periods of time, and out of those fields the asteroseismic measurements are limited to stars that pulsate. Deriving the distances and masses for a larger number of stars than current Gaia observations and asteroseismic measurements can provide is clearly valuable for expanding research into Galactic formation and evolution.
 
Previous work has used spectroscopic parallax to determine distances, e.g. \citealt{Queiroz_2018, Leung_2019, Santiago_2015, 10.1093/mnras/sty2776, Hawley_2002,Burnett_2010}. \citet{Queiroz_2018} developed a Bayesian tool, StarHorse, for calculating spectroscopic parallax distances using APOGEE data along with the Gaia parallaxes.  \citet{Leung_2019} developed a different tool that uses machine learning to accomplish the same goal. Their tool, AstroNN, utilizes the spectral data directly to empirically train a model. The large volume of stars with precise Gaia distances makes machine learning an attractive method for calculating spectroscopic parallax distances because it has the advantage of being an empirically driven approach that is not reliant on stellar modeling. Empirically driven methods can capture and utilize small details in the data that stellar models do not model (\citealt{Ness_2015, Ness_2016, Imig_2022}).

An alternative to asteroseismology for the determination of stellar masses is the use of carbon and nitrogen abundance ratios \citep{10.1093/mnras/stv2830}. As stars ascend the red-giant branch, the observed C and N abundances change at the surface of the star. This is due to the convective regions of stars reaching down deeper into the star and dredging up material processed during the main sequence phase. In the CNO cycle, the nitrogen to oxygen reaction is the slowest, so in regions with CNO reactions the nitrogen abundance will rise and the carbon abundance will fall. As the star goes through dredgeup, this processed material with the shifted carbon/nitrogen ratio is mixed into the outer layers of the star. In more massive stars, there is more CNO processing and the dredgeup moves more material, which results in the [C/N] ratio changing more in higher mass stars (\citealt{vincenzo2021cno,2019ApJ...872..137S,refId0,Lagarde_2012, Masseron_2015}). However, using stellar models to predict the mass is problematic due to uncertainty in the exact mixing process that affects the surface abundances, making the use of model predictions of carbon and nitrogen abundances to estimate stellar mass difficult \citep{10.1093/mnras/stv2830}. As a result, empirical calibration based, e.g., on asteroseismic masses, is an attractive option. A potential issue with empirical calibration are the possible primordial variations in the [C/N] ratio \citep{Masseron_2015}. This issue is addressed in section \ref{massmodel}.

In this paper, two different neural networks are used to estimate stellar masses and distances using the APOGEE Stellar Parameters and Abundances Pipeline (ASPCAP) (\citealt{2017AJ....154...94M, 2016AJ....151..144G, Nidever_2015}) stellar parameters as inputs. This is in contrast to the methodology of \citet{Leung_2019} who use the full spectrum as the input. Using a smaller number of input parameters means that the model architecture becomes much simpler, which reduces the computational load and reduces the risk of over fitting. An advantage of our model is that the noise in the full spectrum does not influence the model training. ASPCAP utilizes the whole spectra to obtain the stellar parameters and abundances, so the input parameters of our model do not contain the signal noise found in the spectrum. Another advantage of using the parameters instead of the full spectrum is that the model is not intrinsically tied to data from APOGEE, so our model could use stellar parameters derived from other data sources. Finally, training on parameters allows us to present empirical relations between parameters and masses that may be of use in constraining stellar mixing models. 

Our models are empirically trained using accurate distance and mass data. For the distance model the training data come from Gaia targets with low parallax uncertainty, while for the mass model mass estimates come from asteroseismic measurements from the APOKASC 3 catalog (Pinsonneault et al. in preparation). The mass model is to some extent an expansion of the results presented in \citet{10.1093/mnras/stv2830} with an expanded data set and using a neural network instead of polynomial relations between masses, stellar parameters, and abundances. Our work is also similar to \citet{10.1093/mnras/sty2776}, with the main distinctions being the training on a much larger sample of stars from new catalogs (Gaia DR3, APOKASC 3) and the application of the models to all stars for distance predictions, and all evolved stars for age predictions.

We also present age estimations for giant stars based on the mass predictions from our model, and. This is possible for evolved stars because stellar mass and main-sequence lifespan are tightly correlated. Evolved stars exist for only the final 1-10\% of the stars lifespan so we can infer the age of an evolved star using the predicted stellar mass and isochrones. However, ages can only be inferred in this way for evolved stars. We also provide some extended discussion of different training data for the masses.

Section \ref{sec:setcreation} describes all the various data and data sources used for this project. Section \ref{MLback} covers the machine learning model and how training and testing the model was performed. Section \ref{distmodel} presents the distance model, including the definition of the training set and validation tests. Section \ref{massmodel} presents the mass model. Section \ref{agepred} covers determining stellar age from the mass predictions, and Section \ref{conc} summarizes the results, including limitations. 

\section{Data}\label{sec:setcreation}
Our models use the stellar parameters and abundances from the APOGEE ASPCAP (\citealt{2017AJ....154...94M, 2016AJ....151..144G, Nidever_2015,2017AJ....154..198Z}) pipeline. For this project, the calibrated effective temperature, surface gravity, metallicity ([Fe/H]), carbon([C/Fe]), and nitrogen([N/Fe]) abundance parameters were used. These data come from APOGEE DR17 \citep{2022ApJS..259...35A}. Spectra from APOGEE are obtained via a multi-object spectrograph \citep{2019PASP..131e5001W} attached to the Sloan 2.5-meter telescope \citep{2006AJ....131.2332G} at the Apache Point Observatory for the northern hemisphere and the Irenee du Pont Telescope \citep{Bowen:73}.

For estimating distances, the model estimates an absolute magnitude, which is then used with the apparent magnitude and an extinction estimation to derive a distance. For the apparent magnitude we used the 2MASS K-band magnitude. Extinction estimations of stars at lower galactic latitudes ($b<16^\circ$) are based on the RJCE method \citep{2011ApJ...739...25M} using IRAC photometry \citep{2004ApJS..154...10F} when available, and WISE \citep{Wright_2010} for all other stars. Data from IRAC is preferred because of the higher spatial resolution, especially in the crowded low galactic latitude fields. At high galactic latitudes stars are more likely to be behind dust, so the extinctions from \citet{1998ApJ...500..525S} are used. It should be noted that any systematic errors in the extinctions will lead to systematic errors in the distance predictions.

The training labels used for the distance model came from absolute magnitudes derived from Gaia parallax data in DR3 \citep{2016A&A...595A...1G}. Specifically, we used distance estimates made using a weak distance prior, which improves the distance estimate over that obtained by simply inverting the parallax \citep{Bailer_Jones_2018}.

The data for mass estimation comes from APOKASC (Pinsonneault et al. in preparation), which is an asteroseismic and spectroscopic survey combining data from APOGEE and the Kepler Asteroseismology Science Consortium (KASC). This is supplemented with additional mass data for sub-giants (using asteroseismology) and main sequence stars (using isochrones) from \citet{2013MNRAS.429.3645S} and \citet{2020}, respectively.

The data sets were trimmed based on several criteria to eliminate bad data from model training. For both models, stars with missing data values for any of the derived parameters were excluded, along with stars flagged with the STAR\_BAD bit. Additional cuts and additions to the input data are discussed in following sections for the distance and mass neural nets since those cuts are specific to their respective neural nets. 

\section{Machine learning background}\label{MLback}

For this work, two small supervised regression models were trained to predict the absolute magnitude and masses of stars, respectively, using the parameters log g, effective temperature, metallicity and the carbon and nitrogen abundances. A supervised machine learning model is one in which independent data is used to train and grade the predictions of the model. Model training iterates over the training data while changing the internal model parameters with the goal of reducing the value of the loss function, which quantifies the performance of the model.

The data were split into a training, validation, and a test set. The training set comprised of 64\% of the training stars. The rest of the stars are part of the validation set and the test set, comprising 16\% and 20\% of the data respectively. The test set is withheld from training and used to determine the model's performance after training is complete. The validation set is also withheld during model training like the test set, but it is used during training as an unbiased estimate of the model's performance while tuning the model parameters. The sizes of the training and validation set were chosen based on the 80/20 rule, where 20\% of the data is used for testing while the remaining 80\% is used in training/validation \citep{tennenholtz2018train}. If the model is over-fitting the training data, then the loss function of the validation set will change little as the training set loss function decreases. Since the validation set has an indirect influence on model training, it also can not be fully trusted for testing. After training the final performance of the trained model is checked using the test set. It is used to verify the model is making reliable predictions and to find issues such as over/under-fitting not addressed in training.

    \subsection{Machine learning architecture}\label{MLarch}
    
    To train and apply the models, we used the Tensorflow Python package. Tensorflow is an open source library that makes the creation and training of neural networks with many layers relatively easy \citep{tensorflow2015-whitepaper}. The model architectures used in both models was a simple single dense layer with 8 neurons that feed into an single neuron output layer. This architecture was chosen after iterating over several model architectures and examine performance with the test set. We found that larger models tended to have over-fitting and smaller ones do not capture the information it needs from the input data. For the loss function we used the mean squared error normalized by the uncertainties.
    
    The input parameters were normalized and scaled before training by subtracting the mean of each parameter and dividing by the standard deviation. This process reduces the overall model training time. An important note in doing this is that any data used as input for the models after training will have to be normalized in the same way. The models were then trained until the loss function of the validation set stopped decreasing, which occurred at around 80 iterations for our models (each epoch meaning a complete run through of the entire data set). 

\section{Distances}\label{distmodel}

    \subsection{Training set} \label{trainingset}
    
\begin{figure}

\begin{tabular}{cc}
  \includegraphics[width=75mm]{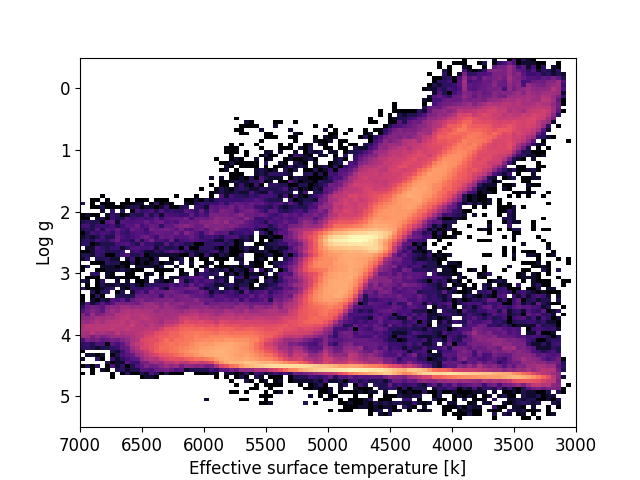} &   \includegraphics[width=75mm]{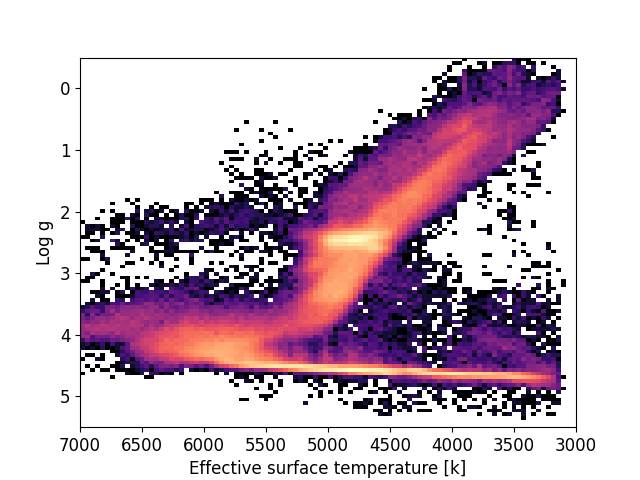} \\
(a) Full data set kiel diagram  & (b) Post uncertainty cut kiel diagram \\[6pt]
 \includegraphics[width=75mm]{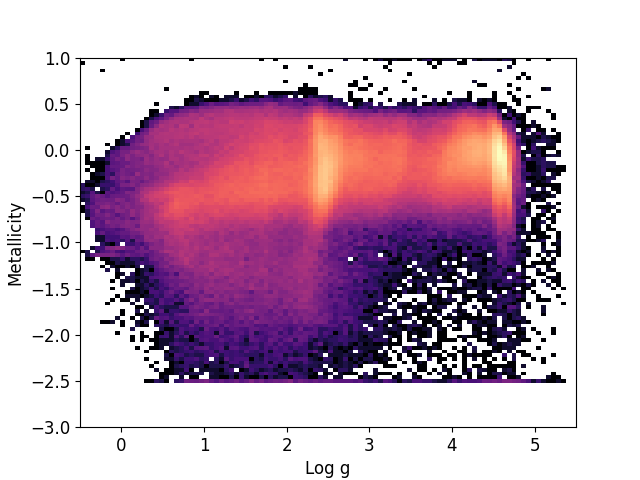} &   \includegraphics[width=75mm]{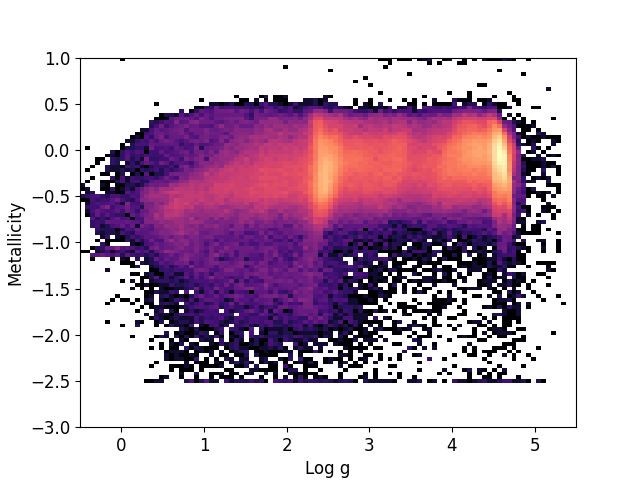} \\
(c) Full data set log g vs metallicity & (d) Post uncertainty cut log g vs metallicity\\[6pt]
\end{tabular}
\caption{Density plots showing the kiel and metallicity diagrams and  for the full data set (a and c) and for the training set after the 15\% distance uncertainty cutoff (b and d). Color scale is logarithmic.}
\label{fig:dataCoverage}

\end{figure}

    For training we adopted all stars with Gaia parallax uncertainties less than 15\%. These stars tend to be within 2-5 kpc, depending on luminosity. We adopted an upper cutoff of 15\% because it reduced the influence of stars with poor distance precision while retaining a large number of stars covering the input parameter space. A smaller cutoff would have resulted in a smaller input parameter space covered by the training set. We also adopted an extinction cutoff of $A_K < 0.5$. This prevents stars with large extinctions from influencing training. The large extinction stars have higher extinction uncertainties which results in larger absolute magnitude uncertainties, which is a problem for our model which uses the absolute magnitude as the training labels. 
    The machine learning model can only be applied over the region of parameter space where the stars in the training set are located. Unfortunately, cutting stars with higher Gaia uncertainty results in parts of the training parameter space having insufficient numbers of stars for training. Figure \ref{fig:dataCoverage} show the parameter space coverage of the full data set (a and c) and the training data set (b and d) after removing high uncertainty stars and randomly sampling the full data set to create the training set as discussed in section \ref{MLarch}. Figure \ref{fig:dataCoverage}b shows there is sparser coverage along the upper giant branch, especially at log g $<$ 1. Figure \ref{fig:dataCoverage}d shows the same lack of stars at low surface gravity and also that for stars with $0.5 < $ log g $< 2.5$ there is a significant drop in the number of stars with [Fe/H] $<$ -1. These cutoffs result in 487543 stars remaining (0.66\% of DR17) in the training set.
    
\begin{figure}
    \gridline{\fig{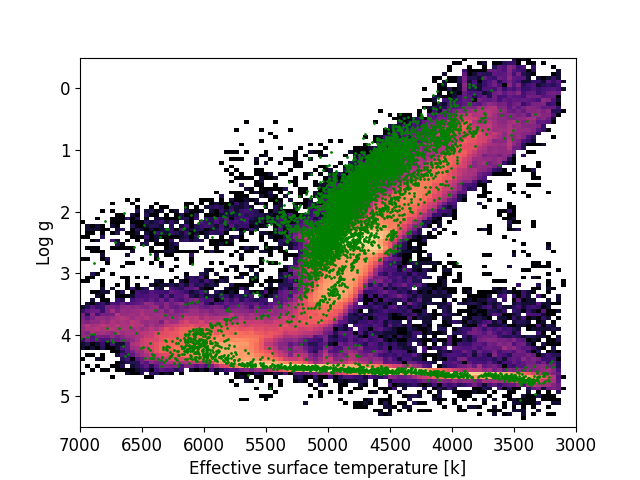}{0.45\textwidth}{(a) kiel Diagram of training space after addition of cluster stars}
    \fig{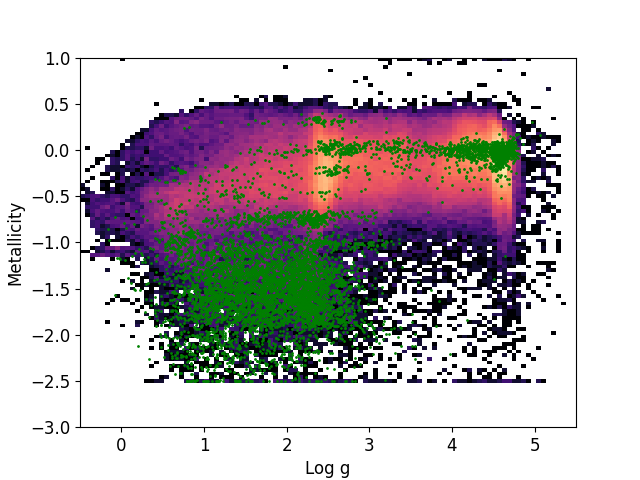}{0.45\textwidth}{(b) Log g vs Metallicity of training space after addition of cluster stars}}
        \caption{HR diagram showing stars with added cluster distances shown in green(a) Coverage of log g and metallicity parameter space showing stars with added cluster distances shown in green (b). }
    \label{fig:ClusterStars}
\end{figure}

    To increase the parameter space covered by the training data set, especially in the low metallicity regime, stellar distances were augmented with distance data for star clusters, for which independent distances are available from main sequence fitting or other methods. A list of all the clusters used is given in Table \ref{table:clusterTable}.  This resulted in the addition of 7039 stars with cluster distances that were originally removed due to high parallax uncertainty.  Figure \ref{fig:ClusterStars} shows the increase in parameter space coverage that comes from including cluster stars. The typical uncertainties for these cluster distances are between 7\% - 10\% (\citealt{2010AJ....139..476P, G_van_de_Ven_2005, Shao_2019, Friel_2010}). The addition of these stars is useful for training our model since they are mostly metal-poor stars, thus they help fill in a region of parameter space that previously had insufficient numbers of stars. 

% \begin{table}
%     \centering
%         \begin{tabular}{ |p{4cm}||p{5cm}| }
%          \hline
%          \multicolumn{2}{|c|}{Cluster List} \\
%          \hline
%          Cluster & Adopted distance [Kpc] \\
%          \hline\hline
%          M92   & 8.2\\
%          M15 &   10 \\
%          N5466 & 15.9\\
%          N4147    & 19   \\
%          M2 &   12  \\
%          M13 & 6.8  \\
%          M3 & 10   \\
%          M5   & 8   \\
%          M12 &   5   \\
%          M107 &6  \\
%          N2243    &4.5    \\
%          N2158 &   4   \\
%          N2420 & 2.5   \\
%          N188 & 1.7   \\
%          M67   & 0.9   \\
%          N7789 &   2.2   \\
%          N6791 &4.078   \\
%          N5053    &17.4 \\
%          M68 &   10 \\
%          N6397 & 2   \\
%          M55 &5  \\
%          M22   & 3   \\
%          M79 &   13  \\
%          N3201 &5  \\
%          M10    &4  \\
%          N6752 &   4 \\
%          Omegacen & 5 \\
%          M54 & 27   \\
%          Pal5 & 23   \\
%          N6544 & 2.7   \\
%          N288 & 8.8 \\
%          N362   & 9   \\
%          N1851 &   12    \\
%          M4 & 2   \\
%          N2808    & 9.6   \\
%          47TUC &   5    \\
%          N6388 & 10   \\
%          N6553 & 6.01  \\
%          \hline
%         \end{tabular}
%     \caption{List of star clusters used to supplement the distances.}

%     \label{table:clusterTable}
% \end{table} 

\begin{deluxetable}{cc}
\tablecaption{List of star clusters used to supplement the distances.\label{table:clusterTable}}
\tablehead{
\colhead{Cluster} & \colhead{Adopted distance [Kpc]}
}
\startdata
 M92   & 8.2\\
 M15 &   10 \\
 N5466 & 15.9\\
 N4147    & 19   \\
 M2 &   12  \\
 M13 & 6.8  \\
 M3 & 10   \\
 M5   & 8   \\
 M12 &   5   \\
 M107 &6  \\
 N2243    &4.5    \\
 N2158 &   4   \\
 N2420 & 2.5   \\
 N188 & 1.7   \\
 M67   & 0.9   \\
 N7789 &   2.2   \\
 N6791 &4.078   \\
 N5053    &17.4 \\
 M68 &   10 \\
 N6397 & 2   \\
 M55 &5  \\
 M22   & 3   \\
 M79 &   13  \\
 N3201 &5  \\
 M10    &4  \\
 N6752 &   4 \\
 Omegacen & 5 \\
 M54 & 27   \\
 Pal5 & 23   \\
 N6544 & 2.7   \\
 N288 & 8.8 \\
 N362   & 9   \\
 N1851 &   12    \\
 M4 & 2   \\
 N2808    & 9.6   \\
 47TUC &   5    \\
 N6388 & 10   \\
 N6553 & 6.01  \\
\enddata

\end{deluxetable}

\subsubsection{Binary removal}

\begin{figure}
    \gridline{\fig{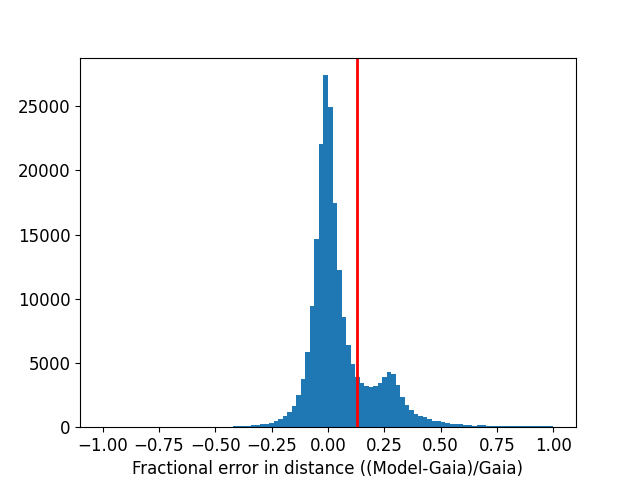}{0.45\textwidth}{(a) Fractional error of dwarf stars including binaries}
    \fig{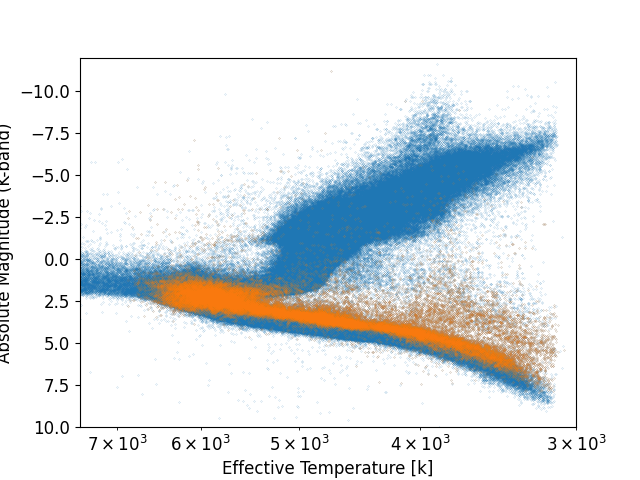}{0.45\textwidth}{(b) kiel diagram showing where the removed stars exist}}
        \caption{(a) Residual plot showing dwarf stars over Gaia distance. Red line indicates the 0.13 cutoff threshold (b) kiel diagram showing where binary stars live}
    \label{fig:BinaryCut}
\end{figure}

Since the distances are calculated from the apparent magnitude and the model-predicted absolute magnitude, there will be systematic errors for unresolved binaries that will be brighter than their derived parameters would suggest. This situation is further complicated for double-lined spectroscopic binaries where the derived parameters may be less accurate if the light from multiple stars contributes to the spectrum. This in turn might cause predictions for all stars to have systematic errors because the errors from binary stars will affect the model training. To remove binaries from the training set, we trained and ran a distance model and compared the distances to Gaia distances. We attempted to just use the binary flag and radial velocity scatter from DR17, but these did not remove enough binaries from the training set.  Stars in binaries will have a higher residual due to the model predicting a dimmer luminosity for them. This results in a bi-modal distribution in fractional error. Stars with a residual fractional distance above 0.13 are then removed from the training set. The threshold was determined by picking the residual value that splits the bimodel distribution so that the high residual samples are removed. Figure \ref{fig:BinaryCut} shows the residual plot with the cutoff line (a) and where the cut stars are located on an HR diagram compared to the rest of the data set. This process improved the performance of the model but does also cut out some non-binary stars. Other methods for binary identification such as radial velocity scatter or eclipsing binaries do not have as high a false positive rate but they do not remove enough of the binary stars from the training set to improve the model. It was important to remove as many binaries from the training set as possible, so the high false positive rate of our method was deemed acceptable. This method removed 134092 stars from the training and test sets, leaving a final training set size of 382673 stars (52\% of DR17) and a test set size of 95668 stars (13\% of DR17). Note that in the full data sample, the model is applied to the binary stars and those will still have systematic errors. However, the systematic errors mostly affect dwarfs, for which Gaia distances are likely to be more reliable than our model. For most star systems with a giant star the difference in luminosity is much greater than between dwarf companions, so the contamination of the companion star around a giant is less significant.
   
\subsection{NMSU isochrone-based distances}\label{nmsudist}

To compare different techniques and also to document some previous work, we present a separate set of distances determined using observed stellar parameters and isochrones, hereafter referred to as NMSU distances. Distances derived using this technique have been distributed in a previous SDSS Value Added Catalog\footnote{\url{https://www.sdss.org/dr14/data\_access/value-added-catalogs/?vac\_id=apogee-dr14-based-distance-estimations}}
and were used in \citet{Weinberg2019}.  For these distances, we use PARSEC isochrones \citep{Bressan2012} with a \citet{Kroupa2001} IMF to identify the expected relative number of stars as a function of effective temperature, surface gravity, metallicity, and absolute magnitude, under the assumption of a constant star formation history. We multiply this by the likelihood of the observed stellar parameters for each star, assuming normally distributed uncertainties given by the DR17 parameter uncertainties, and marginalizing to get the probability distribution function of absolute magnitude. This is then combined with our extinction estimates and apparent magnitudes to provide a probability distribution function in distance. In subsequent discussion, we use the median of the distance probability distribution function to compare with our neural network distances.
    
\subsection{Verification}

\begin{figure*}[ht]
        \centering
        \includegraphics[width=\textwidth]{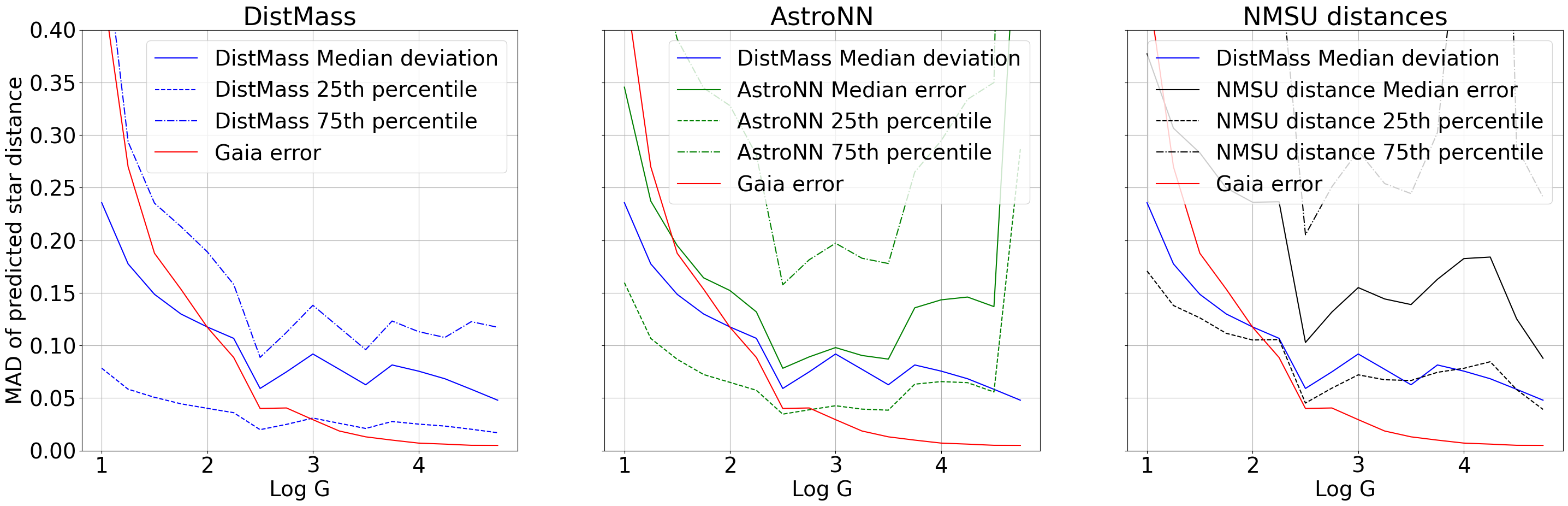}
        \caption{Blue line in all plots shows the MAD of test stars versus surface gravity for our model. Red shows the Gaia parallax uncertainty. Figure a shows just our model compared to Gaia error with the dashed lines showing the quartile of the absolute deviation. Fig b shows AstroNN's \protect\citep{Leung_2019} MAD and quartile ranges in green compared to Gaia and our model. Fig c shows the Bayesian NMSU distance MAD in black. The deviations reported here are the listed stellar distance from each method compared against the Gaia distances}.
        \label{fig:MAD_Dist}
    \end{figure*}

Figure \ref{fig:MAD_Dist}a shows the median absolute deviation (MAD, solid blue line) converted to standard deviation and the percentiles of the absolute deviation range (dashed blue lines) of our distances compared to the Gaia distances, as a function of surface gravity. This difference is largest at lower surface gravities, which we attribute partly to the poorer coverage in this region of our training set, but also to the larger uncertainties in the Gaia distances to these stars, which tend to be at larger distances. At higher surface gravities, the model achieved a median deviation of less than 10\%; of course, for these stars, which are generally close, the Gaia distances are significantly more accurate.

Figures \ref{fig:MAD_Dist}b and \ref{fig:MAD_Dist}c show the same comparisons for AstroNN \citep{Leung_2019} and NMSU distances. AstroNN recommends using their distances that are weighted with Gaia distances, but here we compare their unweighted distance predictions since otherwise we would be comparing to Gaia for nearby stars. Figures \ref{fig:MAD_Dist}b \& c shows that our neural network (in blue) tends to have more accurate distance predictions than AstroNN and NMSU distances across all log g values. This may be due to AstroNN being influenced by noise in the full spectra thus reducing its accuracy while the NMSU distances are subject to uncertainty in the isochrones and the chosen IMF.

\begin{figure*}[ht!]
    \centering
    \includegraphics[width=\textwidth]{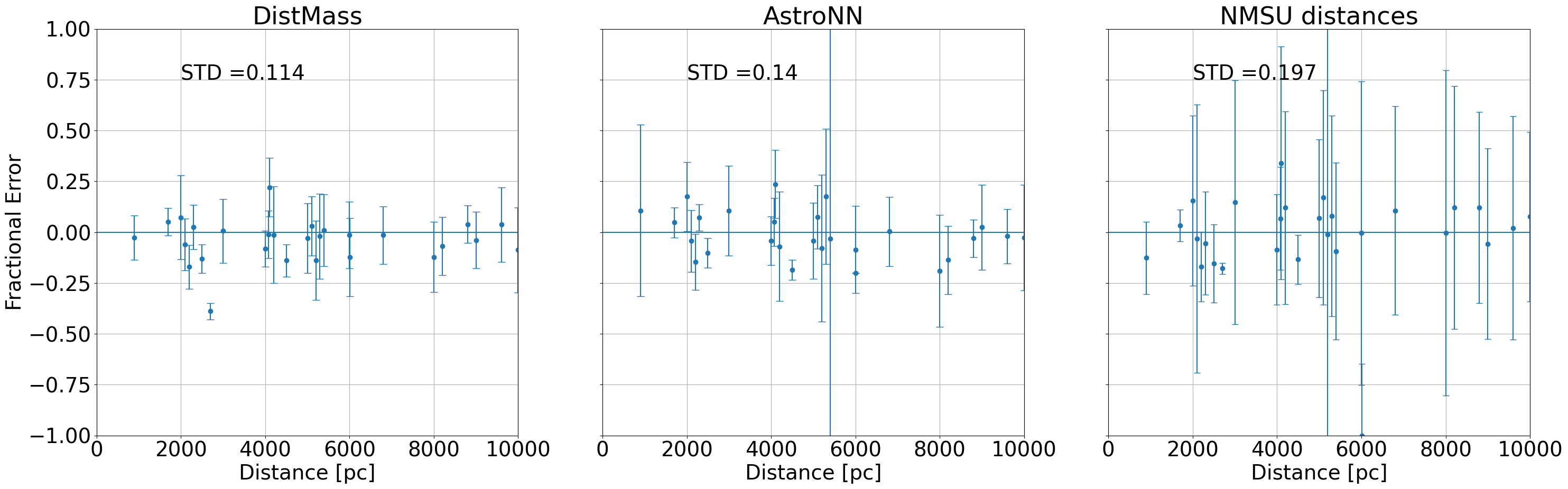}
     \caption{Model fractional error vs cluster distance for cluster stars. Blue points shows average predicted distance to each cluster. The error bars show the standard deviation of the scatter in distance predictions for stars in each cluster around the mean distance. Scatter of the cluster distance averages is printed for each distance method. Figure a shows our model results, figure b shows AstroNN \protect\citep{Leung_2019}, and figure c shows NMSU distances. STD of the cluster distance average (orange points) is printed for each plot}
     \label{fig:Clusterdistances}
 \end{figure*}

Additional validation comes from distances to cluster stars, both from the scatter in distances to cluster members and from comparison to literature distances. Figure \ref{fig:Clusterdistances}a shows the performance of the neural network on stars in the clusters given in Table \ref{table:clusterTable}. The predicted distances for each star in each cluster were averaged to give estimated cluster distances shown with the blue points in figure \ref{fig:Clusterdistances}a. The error bars show the standard deviation of the scatter in distance predictions for stars in each cluster around the mean distance. The median fractional error across the cluster mean distances is -0.06 and the standard deviation of the fractional error in cluster mean distances is 0.114. Model performance was also judged by examining the median scatter of predicted star distances around the true cluster distance for each cluster; we find a median standard deviation of 0.14 in fractional distance .  

Figures \ref{fig:Clusterdistances}b and \ref{fig:Clusterdistances}c show the same comparison for AstroNN and NMSU distances. The median fractional error of mean cluster distances are closer to zero than our model, 0.038 for AstroNN and -0.042 for NMSU, but the standard deviation of mean cluster distances is  higher at 0.14 and 0.197 respectively. The median standard deviation within clusters for the comparison models is 0.37 and  0.4 for AstroNN and NMSU, respectively. 

Uncertainties for our distance predictions come from a quadratic fit of the median fractional error vs log g. We do not have any inclusion of the input uncertainties. We recommend using the distances from our model for stars with larger Gaia parallax uncertainty, but for stars with small Gaia parallax uncertainty, the Gaia distances are preferred. Our catalog contains a column with our model distances weighted with the Gaia distances by the quoted uncertainties in each measurement. 

\section{Masses}\label{massmodel}

     A second goal of this paper is to develop a model that estimates the mass (and age) of stars using the stellar parameters from APOGEE DR17.

\begin{figure}[h!]
    \gridline{\fig{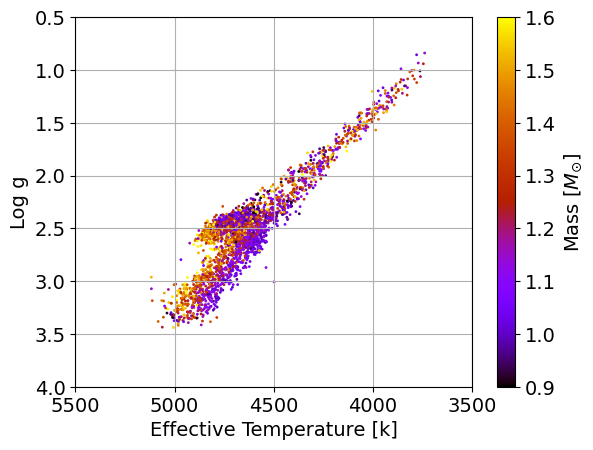}{0.45\textwidth}{(a)Asteroseismic masses, $[Fe/H]=0 \pm 0.025$}
    \fig{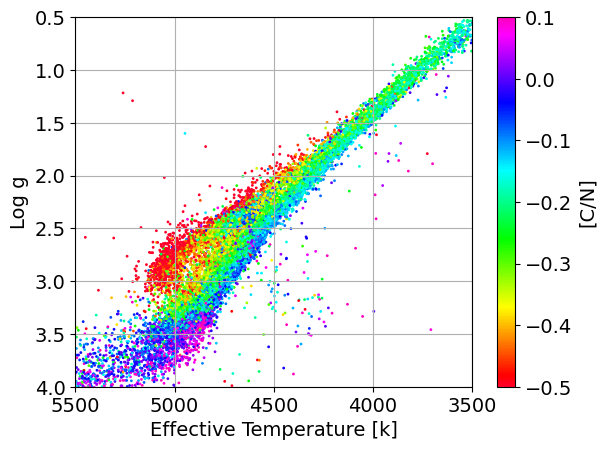}{0.45\textwidth}{(b) [C/N] ratio, $[Fe/H]=0 \pm 0.025$}}
        \caption{(a) kiel diagram of APOKASC stars $<1.6\: M_\odot$ with $-0.05 < [Fe/H] < 0.05$ and color coded by asteroseismic mass. (b) kiel diagram showing the sub-giant and  red giant branch for stars with $-0.025 < [Fe/H] < 0.025$ color coded by [C/N].}
    \label{fig:hrEEP}
    \end{figure}

     Figure \ref{fig:hrEEP}a shows a kiel diagram for stars from the APOKASC catalog (Pinsonneault et al. in preparation) with solar [Fe/H] and color coded by mass which demonstrates that, especially along the upper giant branch, stars of different masses cannot be distinguished by Teff and log g alone. Fortunately, the [C/N] ratio can be used to separate stars of different masses \citep{10.1093/mnras/stv2830}. As a star evolves onto the red giant branch, the first dredge-up results in a drop of the [C/N] ratio, where higher mass stars have a larger change. Figure \ref{fig:hrEEP}b shows a kiel diagram of solar [Fe/H] stars from APOGEE DR17 color coded by [C/N] where the dredge-up mixing process is apparent near log g = 3.5.

% \subsection{Strength of the [C/N] - mass relation}

\begin{figure}
        \centering
        \includegraphics[width=0.95\textwidth]{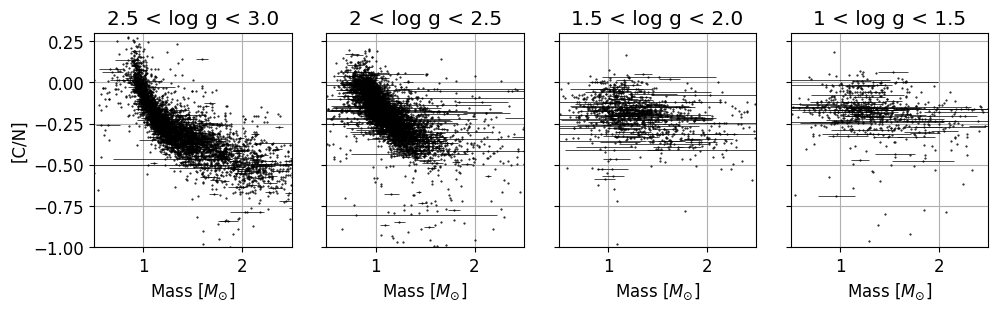}
        \caption{[C/N] - mass relation binned by log g. The masses shown here are the corrected Mosser masses from APOKASC. Every 10 points has an error bar for mass}
        \label{fig:CNMASSRel}
    \end{figure}

    Mass/age determination for evolved stars using [C/N] requires a correlation between the measured [C/N] and stellar mass across parameter space. Figure \ref{fig:CNMASSRel} shows the relation between [C/N] and APOKASC mass along with some representative error bars binned by log g. At higher log g the [C/N] - mass relation is tight and defined, however, at low log g the relation is less definitive. There is little physical reason for the [C/N] - mass relation to change at low log g for most stars, although extra-mixing may need to be considered for metal poor stars ([Fe/H]$<-0.5$)(\citealt{2005essp.book.....S, 2019ApJ...872..137S}). The error bars shown in Figure \ref{fig:CNMASSRel} grow larger at lower log g values, with an average mass uncertainty reaching up to $0.3 M_{\odot}$. Therefore, the explanation for the less clearly defined [C/N] - mass relation at lower log g is the increased uncertainty in the APOKASC masses.

\begin{figure}[h!]
    \gridline{\fig{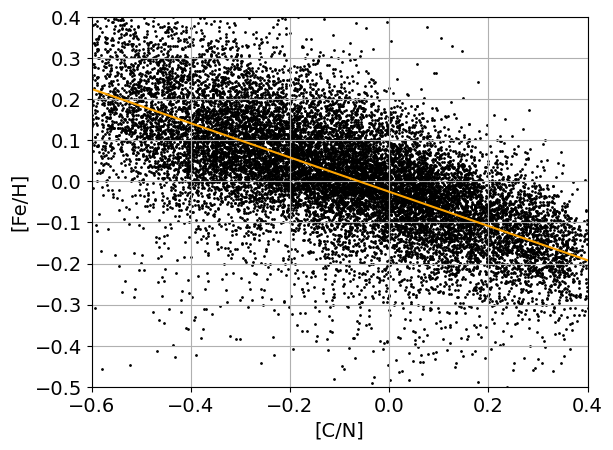}{0.45\textwidth}{(a) Sub-giant stars from APOGEE DR17 $4 < log\, g < 3.5$ and $5500K < T_{eff} < 4750K$ }
    \fig{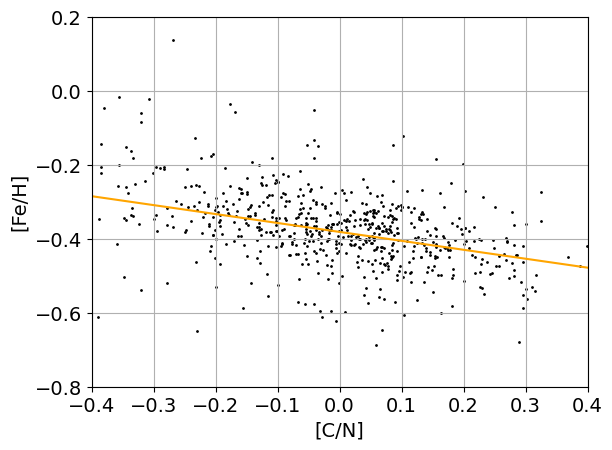}{0.45\textwidth}{(b) RGB stars from APOKASC $3 < log\, g < 2.5$ and $1.5 < M_{\odot} < 1.75$}}
        \caption{(a) [C/N] as a function of [Fe/H] for sub-giant stars. Slope shows the [Fe/H] - birth [C/N] relation (b) [C/N] as a function of [Fe/H] for RGB stars. Star selection was limited to a narrow mass range in order to highlight the [C/N] - [Fe/H] relation.}
    \label{fig:FeHCN}
    \end{figure}

% \subsection{Primordial [C/N] variations}\label{PriCN}
    Figure \ref{fig:hrEEP}b shows a range in the pre first dredge-up [C/N] along the sub-giant branch stars below log g 3.3. The range in pre first dredge-up [C/N] is a result of variations in the birth [C/N] of these stars. The first dredge-up process does not completely erase the original birth [C/N], therefore these variations in the birth [C/N] will influence the observed [C/N] in red giant branch stars and consequently can affect stellar mass/age predictions based on [C/N] abundances if the birth abundances are not taken into account (\citealt{10.1093/mnras/stv2830, vincenzo2021cno}; Roberts et al,in preparation). \citet{vincenzo2021cno} and Roberts et al (in preparation) observed that birth [C/N] correlates with [Fe/H] and [Mg/Fe]. Figure \ref{fig:FeHCN}a shows the [C/N] - [Fe/H] relation for sub-giant stars, which means the measurement is of birth [C/N]. Figure \ref{fig:FeHCN}b also shows [C/N] and [Fe/H] but for APOKASC RGB stars in a narrow mass bin, which indicates that the [Fe/H] - [C/N] relation is important after first dredge-up. This clearly shows that including [Fe/H] in chemical mass/age is important. 
    While Roberts et al (in preparation) also suggest [Mg/Fe] correlates with birth [C/N], the additional use of [Mg/Fe] would be problematic for our mass/age estimations and future use cases of our estimations. [Mg/Fe] has been shown to have a correlation with age (\citealt{Haywood2013, Bensby2014, 2012A&A...542A..84D}), however the [Mg/Fe] - age  correlation has been found to vary across the Galactic disk (\citealt{2018MNRAS.475.5487S, 2018MNRAS.477.2326F, 2021A&A...655A.111K, 2022A&A...660A.135V, Imig_2023}). The spatial variation in the [Mg/Fe] - age  correlation could cause any chemical stellar mass/age prediction model to spatial systematics if [Mg/Fe] was included in the model. We also intend our mass/age predictions to be useful for mapping the [Mg/Fe] - age correlation to better constrain Galactic chemical evolution models. Including [Mg/Fe] in the prediction model would tie the [Mg/Fe] abundances to our mass/age estimations and therefore our estimations could not be used as an independent test of [Mg/Fe] chemical evolution. As a result of this we exclude [Mg/Fe] and include [Fe/H] for our mass/age estimation models.

     In order to leverage the [C/N] - mass relationship and estimate the masses of evolved stars, we have built a second neural network model that uses the effective temperature, surface gravity, metallicity, carbon, and nitrogen abundances as input parameters. The training data is constructed from asterosiesmic masses in the APOKASC catalog which uses data from Kepler observations as described in Section \ref{sec:intro}.
    
\subsection{Spatial Variations in the [C/N] - mass relation}

\begin{figure}
    \centering
    \includegraphics[width=1\linewidth]{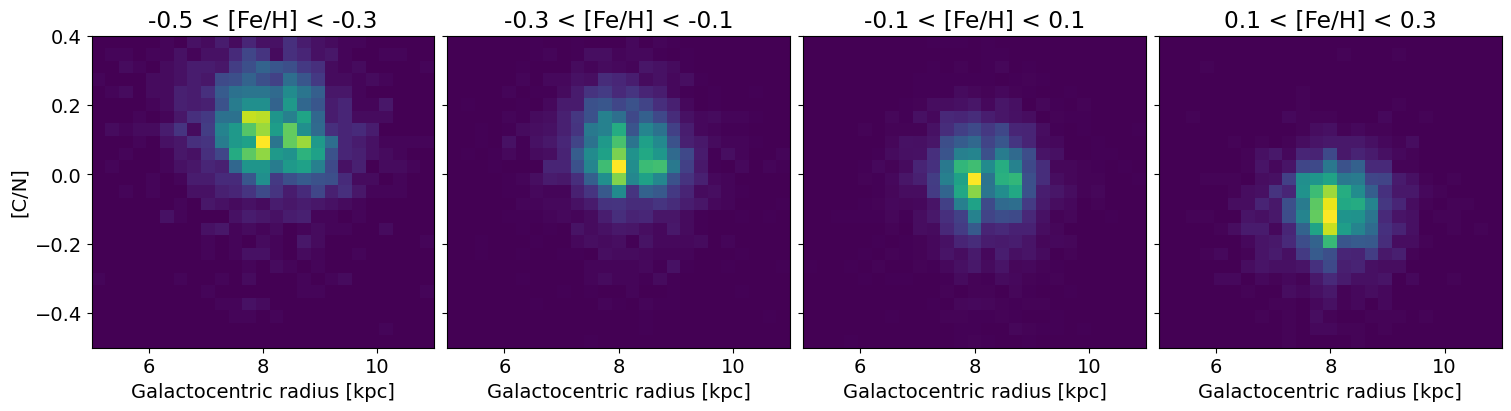}
    \caption{Histograms of [C/N] vs Galactocentric radius for pre-dredge up subgiants binned by [Fe/H]. The location of the
    distribution is different for each [Fe/H] bin due to the [C/N] - [Fe/H] relation. There is no  apparent trend in the [C/N] - [Fe/H]
    relation over the examined range. However, the range we can examine the [C/N] variation is limited  to 2 kpc by the lower
    luminosity of subgiants}
    \label{fig:CNRadius}
\end{figure}

    Since our goal is to reliably predict stellar masses across a large volume of the Galaxy, we require the training stars to be representative of the entire sample we predict ages for. Our training data is limited to stars in the original Kepler field which does not sample a wide range of Galactocentric distance, although we apply the age model to all of the stars in DR17 which does span a large region of the Galaxy. This requirement is part of the reason we excluded [Mg/Fe] from the mass/age estimation model, since the [Mg/Fe] - age correlation varies across different regions of the Galaxy as described in section \ref{massmodel}. We make the assumption that the training stars are representative of the [C/N]-[Fe/H]-mass relation throughout the Galaxy. However if the [C/N]-[Fe/H]-mass relation also varies as a function of Galactocentric radii then our mass/age estimations will have systematic errors that depend on Galactic location. 
    
    We test our assumption by searching for spatial variations of the [C/N] - [Fe/H] relation in sub-giant stars. If there is little variation, then we can be more confident about our assumption that the Kepler field is representative of the Galaxy. This test can be applied beyond the Kepler field since it does not rely on the asterosiesmology from Kepler. \citet{10.1093/mnras/stv2830} did this test with DR12 data and we have repeated this test using the more extensive DR17 data. Figure \ref{fig:CNRadius} shows a histogram of [C/N] vs Galactocentric radius for all DR17 pre-dredge up stars. We define pre dredge-up stars to have $4 < log\, g < 3.5$ and $5500K < T_{eff} < 4750K$. Figure \ref{fig:CNRadius} does not indicate any spatial trends in [C/N] which is consistent with the results of \citet{10.1093/mnras/stv2830}. Unfortunately, even with the more extensive DR17 data, subgiants are only included within $~ 2kpc$ of the Sun, so our results at larger distances are still subject to potential systematics if the [Fe/H]-[C/N] relation is different at those distances.

    \subsection{[C/N] systematic}
\begin{figure}
        \centering
        \includegraphics[width=0.8\textwidth]{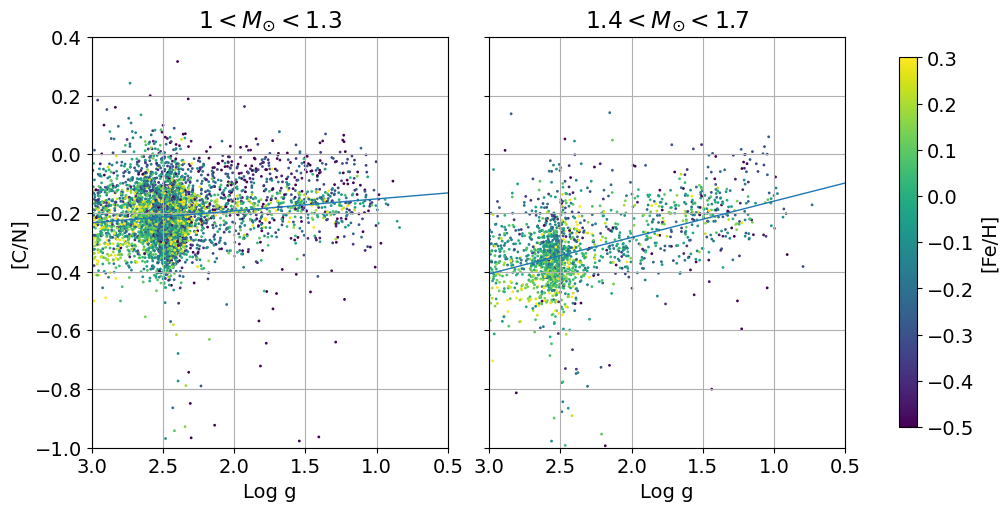}
        \caption{[C/N] vs log g binned by asteroseismic mass. Best fit line shows the positive trend in [C/N]}
        \label{fig:cnLogg}
    \end{figure}

    At low log g, the reported [C/N] from DR17 rises for the stars of the same mass, as shown in Figure \ref{fig:cnLogg}. We expect stars to either keep a constant [C/N] after first dredge-up or a drop in [C/N] from extra mixing. However, Figure \ref{fig:cnLogg} appears to show stars undergoing "unmixing" with a higher [C/N] as they evolve up the RBG, which is nonphysical as discussed by \citet{2022arXiv220408487F}. This appears to be a systematic error in the reported [C/N] that varies with log g. However, in principle, our model trained with this systematic and therefore is unaffected by it. However this highlights that our learned [C/N] - mass relation is tied to the DR17 data and comparisons with models that use non DR17 data must take this into account.
    
    % APOGEE has a systematic error in the reported [C/N] that varies with log g. At low log g, the reported [C/N] rises for the stars of the same mass, as shown in figure \ref{fig:cnLogg}. This appears to show stars undergoing "unmixing" as they evolve up the RBG, which is most certainly nonphysical as discussed by \citet{2022arXiv220408487F}. Our model training does not apply any correction to the [C/N] because the mass predictions rely on the [C/N] - mass relation which is only shifted in [C/N] by the systematic and thus is most likely trained out in the neural network.
    
\subsection{Training space constraints}

\begin{figure}

\begin{tabular}{cc}
  \includegraphics[width=75mm]{Fullspace.png} &   \includegraphics[width=75mm]{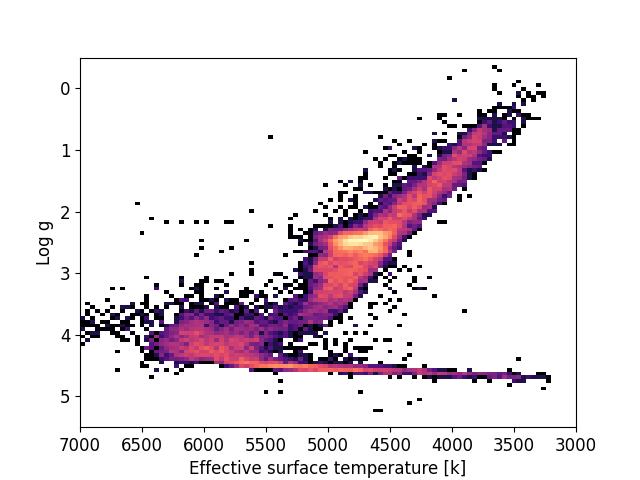} \\
(a) Full data set kiel diagram  & (b) Post mass uncertainty cut kiel diagram \\[6pt]
 \includegraphics[width=75mm]{FullspaceMetal.png} &   \includegraphics[width=75mm]{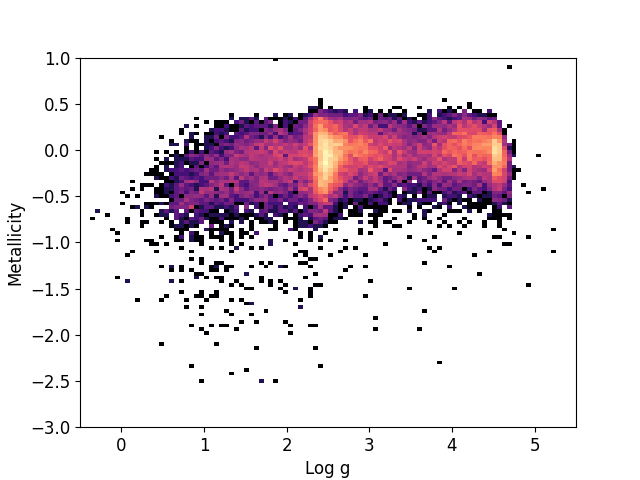} \\
(c) Full data set metallicity & (d) Post mass uncertainty cut metallicity\\[6pt]
\end{tabular}
\caption{Density plots showing the kiel and metallicty diagrams and  for the full data set (a and c) and for the training set after the 10\% mass uncertainty cutoff (b and d). Brighter and colored regions contain higher density of stars}
\label{fig:HRMass}

\end{figure}

The model should not extrapolate too far beyond the space covered by training, otherwise the mass predictions will be impacted by poor model extrapolations. We also want to remove stars with high mass uncertainties from the training set, otherwise they could degrade the model's performance. To achieve this, we chose to cut stars with an APOKASC uncertainty of $>10\%$, resulting in a training set of 13458 stars. These cuts result in significant gaps in parameter space which are shown in Figure \ref{fig:HRMass}. In particular, training data is limited at log g $< 1.5$ and $[Fe/H] < -0.5$, implying that the model will begin to rely of extrapolation for stars at lower log g and more metal poor stars. 

% \subsection{Implementation} \label{massTraining}
    
% For training the mass model, masses from APOKASC (Pinsonneault et al. In preparation) are used, which is an asteroseismic and spectroscopic survey combining data from APOGEE and the Kepler Asteroseismology Science Consortium (KASC). This is supplemented with additional mass data for sub-giants (using asteroseismology) and main sequence stars (using isochrones) from \citet{2013MNRAS.429.3645S} and \citet{2020}, respectively.

% Description of potential APOKASC correction issues go here
\subsection{APOKASC radii corrections}\label{radiiCorr}

The training set for the mass model uses asteroseismic masses from APOKASC 3 (Pinsonneault et al. in preparation). APOKASC 3 contains contains stellar mass estimations using three different seismology results from \citet{Sharma_2016}, \citet{Mosser1}, and \citet{White_2011}. Each of these is included in APOKASC3 with both the raw seismology masses, as well as corrected masses based on a comparison with radii determined by Gaia \citep{Zinn_2019}.

\begin{figure}[h!]
    \gridline{\fig{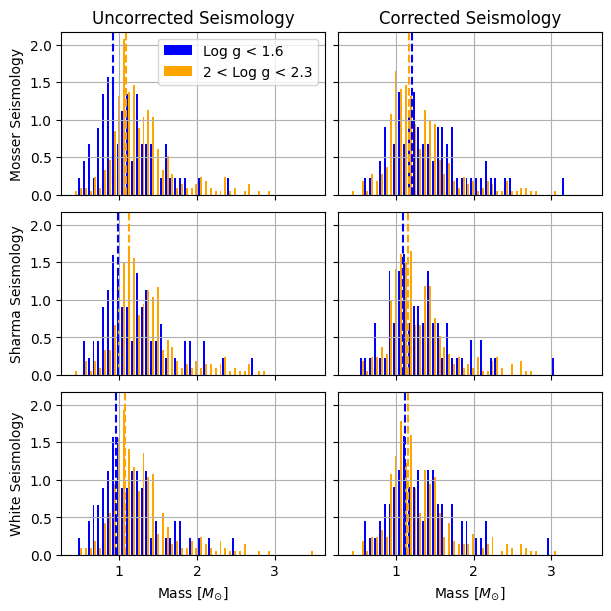}{0.45\textwidth}{(a) $1 < Distance [kpc] <2$}
    \fig{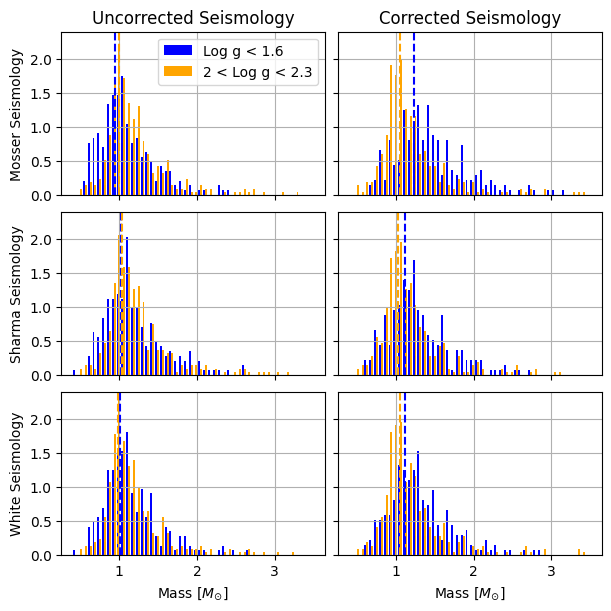}{0.45\textwidth}{(b) $2 < Distance [kpc] <3$}}
        \caption{Mass distribution histograms for the 6 mass data sets in two distance bins. Left columns shows uncorrected masses and the right columns shows the corrected masses. The histograms are split into a low log g bin (blue) and high log g bin (orange).}
    \label{fig:APOMASSHist}
    \end{figure}

To attempt to assess whether there is a preference for one of these data set, we consider mass distributions for stars of different log. We would expect that in a given region of the Galaxy, the distributions of masses for low and high log g stars to be equivalent. In Figure \ref{fig:APOMASSHist} we show normalized histograms of the masses present in each of the six seismology results at 2 different distance bins. Different rows present each of the three seismology results and the columns show the Gaia uncorrected masses (left) and corrected masses (right). The histograms are split into a low log g bin (blue) and high log g bin (orange) and a vertical line shows the mean mass for the log g bins.  In every seismology result, the Gaia corrections shift the low log g stars to higher estimated masses. High log g stars have very small corrections to their masses as expected, because the higher log g stars have very small radii \citep{Zinn_2019}. In most of the results the shifts are fairly mild. In the closer distance bin the corrections bring the mass distributions into alignment, however in the more distance bin the corrections appear to push the distributions out of alignment. in the more distance bin, the corrected \citet{Mosser1} results have the most unequal distributions. So it is not clear if using the corrected seismology results is superior. However, as discussed below, models trained with the corrected results lead to a paucity of low mass stars across the Galaxy.

% Given there are 6 mass data sets to use in APOKASC, and no clear idea on which set is the best representation of actual star masses across parameter space we opted to train a neural network of the same architecture on all 6 data sets. The mass (and age) predictions from these different models are all included in the DistMass catalog.

\subsection{Implementation}

\begin{figure}
        \centering
        \includegraphics[width=0.9\textwidth]{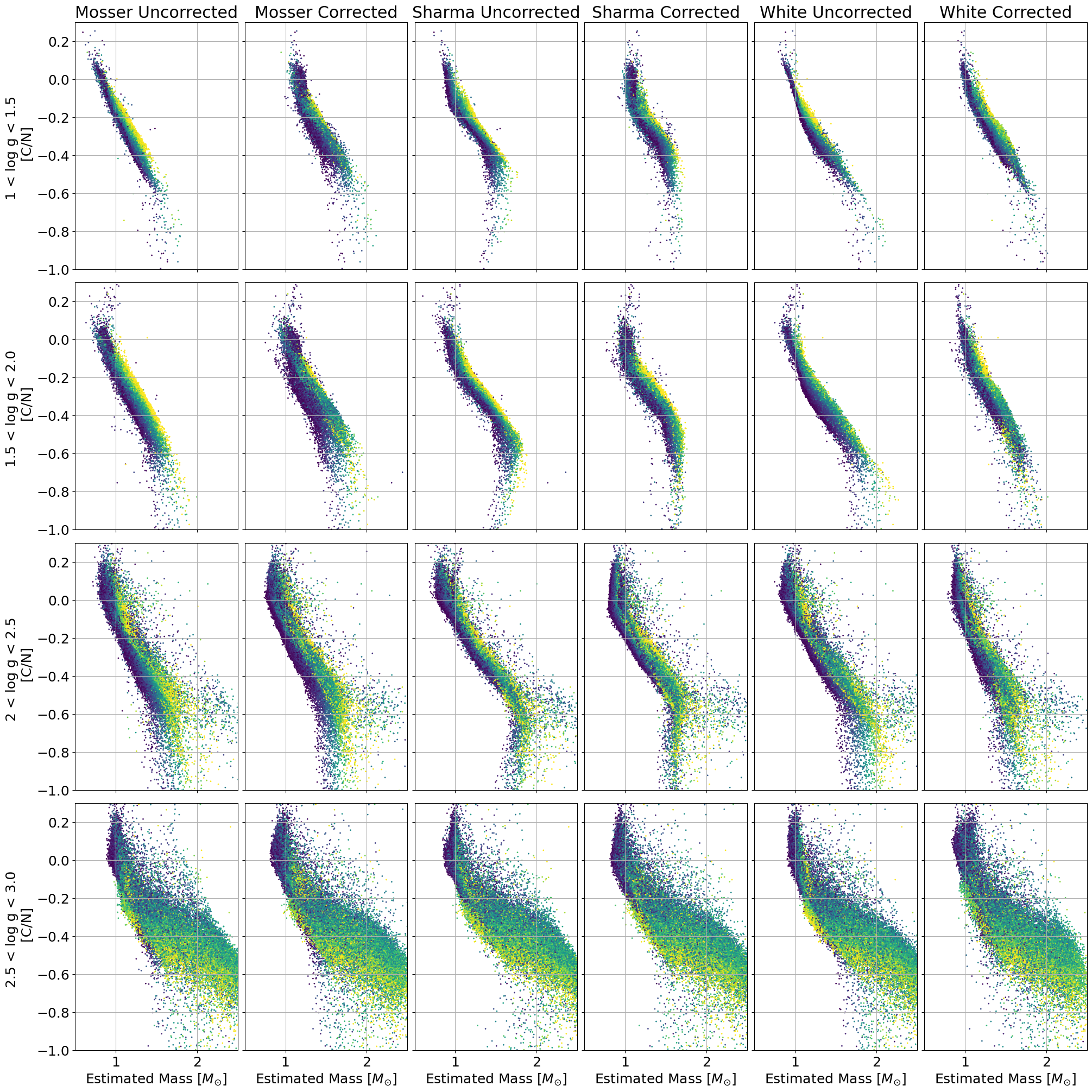}
        \caption{[C/N] - Mass relation for the ML model predictions trained on all six seismologies/corrections. Color coded by [Fe/H] and binned in log g intervals of 0.5.}
        \label{fig:cnMassRelation}
    \end{figure}

In order to better understand the effects of the small differences between the seismologies and corrections we used each as the training set for six independent mass prediction models. Figure \ref{fig:cnMassRelation} shows the derived [C/N] - mass relations for each model, color-coded by [Fe/H] and binned by log g. Each column is a different model trained on the six different seismolgies/corrections and each row is a different log g bin. At high log g there are few differences between the model predictions. However, at low log g the corrections for each seismology appear to shift the [C/N] - mass relation to higher masses. The shift is large enough that there is an absence of low mass stars in the model predictions for the models trained on the Gaia corrected masses. The absence of low mass stars translates to an absence of old stars in the resulting age predictions, which is not an expected feature of the Galaxy nor is it apparent in the training data from Figure \ref{fig:APOMASSHist} which does include low mass stars. The likely cause of this is the Gaia corrections shifting low log g stars to higher masses and the models learning a shifted [C/N] - mass relation.

    \begin{figure*}[ht!]
        \centering
        \includegraphics[width=0.9\textwidth]{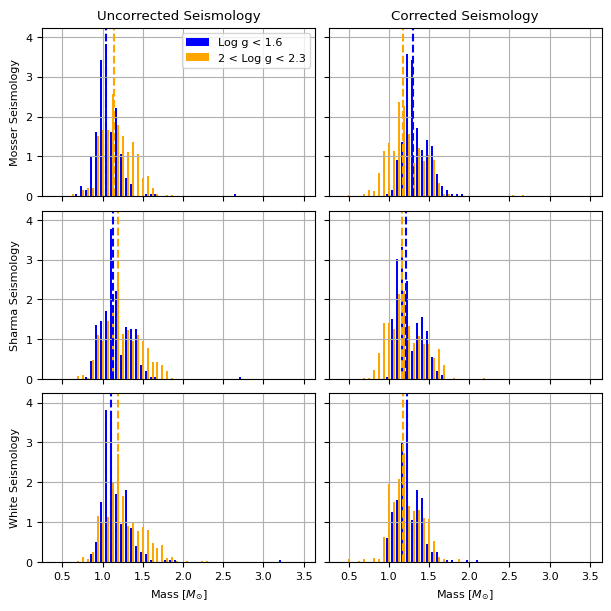}
        \caption{Histograms showing the distribution of masses predicted by the six models. The left column shows uncorrected masses and the right column shows the corrected masses. The histograms are split into a low log g bin (blue) and high log g bin (orange).}
        \label{fig:DistMassHisto}
    \end{figure*}

As discussed in section \ref{radiiCorr}, we expect that in a given region of the Galaxy, the distributions of masses for low and high log g stars to be equivalent. Figure \ref{fig:DistMassHisto} shows histograms of the predicted mass distributions from our six models split into a low log g bin (blue) and high log g bin (orange). The mass distributions from the corrected and uncorrected Mosser models have medians that are offset from each other. The distributions from the Sharma and White models are  equivalent. The shift to higher masses is also apparent for the corrected mass models. In general, the Sharma and White corrected and uncorrected models treat the low log g stars similarly to high log g stars.

Given that it is challenging to know which seismology result is the best one for model training, we elected to produce a model using each set of seismology results and corrections. This resulted in six sets of mass and age predictions in our catalog. However, we note that the mass distributions are more unequal for the Mosser models. We also note that the corrected mass models lack low mass (old age) stars in their predictions. For these reasons \citet{Imig_2023} used the masses and ages from our model trained on uncorrected Sharma masses.

For our reported uncertainties we chose a flat uncertainty of 10\% for all of our predicted masses. This value comes from the median absolute deviation between our predictions and the APOKASC masses for stars at log g $\approx 2.5$. At low log g the median absolute deviation becomes much larger, but this likely results from the fact that the APOKASC uncertainties grow larger at low log g as well. We make the assumption that the uncertainties are similar between high and low log g. This is assumption is based on the similar mass distribution medians shown in Figure \ref{fig:DistMassHisto}. It should be noted, however, that this assumption is weaker for the Mosser models since their distributions are not well matched.

For the ages discussion in the following section, we use the ages from the model using the corrected Mosser seismology since that is what Pinsonneault et al. (in preparation) chose to use for their analysis and this is also the is the seismology used for training the \citet{2019MNRAS.489..176M} stellar ages. However, our final catalog contains masses and derived ages for all six sets of seismology.

% Figure \ref{fig:MAD_mass} shows the median absolute deviation of the model mass predictions from APOKASC masses as a function of log g and stellar mass. The median absolute fractional error is between 0.05 and 0.1, and most stars have $<14.8\%$ mass prediction error. There is a slight increase in median fractional error for low surface gravity stars.

% Uncertainties for our mass predictions come from a quadratic fit of the median fractional error as a function of log g. \textbf{We chose to use a function of log g because the training mass uncertainties are much larger at low log g as seen in Figure \ref{fig:CNMASSRel}}

    % \begin{figure}
    %     \centering
    %     \includegraphics[width=0.5\textwidth]{residualMass.png}
    %     \caption{Fractional error of predicted mass vs the asteroseismic mass color coded by log g.A negative value means the predicted mass was under-predicted, meanwhile, for a positive value, it was over-predicted}
    %     \label{fig:residual_mass}
    % \end{figure}

\section{Ages}\label{agepred}
    \subsection{Age determination from mass}
    For evolved stars there is a relation between the stellar mass and age. Using the predictions from the mass model, we provide age estimations for stars with log g $<$ 3.5 using MIST isochrones  \citep{2016ApJ...823..102C}. The relation between mass and age for evolved stars work because stars only occupy the giant branch for the final 1-10\% of their lives 
    \citep{10.1046/j.1365-8711.1998.01658.x}. The MIST isochrones were used to make a function with initial stellar mass and metallicity as inputs to produce an estimated age. The function assumes all giants of the same mass and [Fe/H] as having the same age. The function covers a mass range of 0.8 to 8 $M_\odot$ and a metallicity range of -0.5 to 0.5 [Fe/H]. Since the function uses the mass estimates from the mass model the age estimation is tied to the performance of the mass model, so the uncertainties for our age estimation are determined by propagating the mass uncertainty through the mass to age function. The metallicities from ASPCAP and the predicted masses from the neural net were used for the age prediction. Due to the nature of the mass to age conversion, the age uncertainties are more sensitive to mass uncertainties at lower masses. So the typical mass uncertainty of $\pm 10 \%$ will result in an age uncertainty of [+0.3, -0.21] log(age) for a $1M_{\odot}$ star, while for a $2M_{\odot}$ star the same mass uncertainty gives an age uncertainty of [0.067, -0.063] log(age).
    
    \subsection{Mass loss} \label{sec: mass}
    As stars ascend into the red giant phase they undergo mass loss. According to the \citet{1977A&A....61..217R} prescription, mass loss is expected to be important primarily for the upper red giant branch due to low surface gravity. Metallicity is also a factor in driving mass loss with higher metallicity stars losing more mass by stellar winds \citep{2001A&A...369..574V}. There also could be mass loss associated with the helium flash that stars under 2.25 solar masses go through (\citealt{1996,2021}); for our sample of stars the most affected region is likely the red clump.
    
    Since the asteroseismic masses used for training are current masses and our neural network is trained only on empirical data, the model does not have any knowledge of stellar mass loss. The mass predictions of the model are estimates of current star mass. If mass loss is significant, then we would expect stars with higher mass loss to have over-predicted ages. 
    
    However, \citet{10.1111/j.1365-2966.2011.19859.x} examined the mass-loss for clusters N6819 and N6791 and show that little appreciable mass-loss is observed in RGB stars with the exception of high luminosity stars at the tip of RGB which would carry over to RC stars as well. It should be noted that these clusters span from solar to super-solar metallicity. Both \citet{10.1111/j.1365-2966.2011.19859.x} and \citet{2017MNRAS.472..979H} examined N6819 mass-loss with asteroseismic data and constrained the possible mass-loss to $<0.03 M_\odot$, which \citet{2017MNRAS.472..979H} noted as almost insensitive to systematics. While the expected mass loss depends on the stellar parameters, these studies suggest that it unlikely to be higher than the constraint of $0.03 M_\odot$. A 1 solar mass star that loses $0.03 M_\odot$ the age prediction will be over-predicted by 11\%; for a 2 solar mass star the over-prediction would be 3\%.
    
    \subsection{Parameter to age relations}

\begin{figure*}[ht!]
        \centering
        \includegraphics[width=0.9\textwidth]{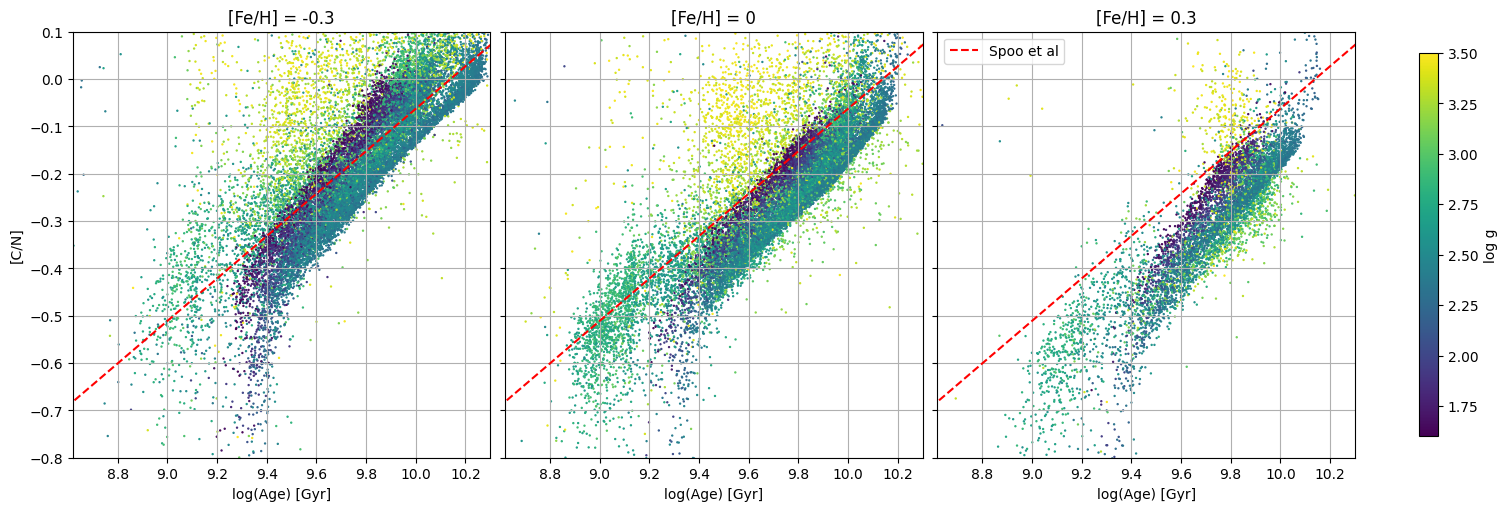}
        \caption{[C/N] vs estimated age binned by [Fe/H] and color coded by log g. Ages are from the model trained on the corrected Mosser seismology. Red dashed line shows the [C/N] age calibration from \citet{Spoo_2022}}
        \label{fig:CNvAge}
    \end{figure*}

    Figure \ref{fig:CNvAge} shows the derived ages as a function of the [C/N] in three narrow [Fe/H] bins color coded by log g. All three show the [C/N] to age relation from \citet{Spoo_2022} in dashed red which was determined by calibrating a fit with the ages of open clusters vs the average [C/N] ratio of a cluster. However, that calibration does not include any metallicity dependence, which is important to the [C/N] - mass/age relation as discussed in section \ref{massmodel} and is apparent across plots in Figure \ref{fig:CNvAge}. The comparison to \citet{Spoo_2022} serves both as another test of our age predictions, and as a demonstration of the importance of including metallicity in any age/mass predictions involving [C/N].
        
    % \begin{figure*}[h!]
    %     \gridline{\fig{ParamAgeSolarMetal.png}{0.4\textwidth}{a}
    %     \fig{ParamAgeAllMetal.png}{0.4\textwidth}{b}}
    %         \caption{C/N vs age for solar metallicity stars (a) C/N vs age for $–0.5 < [Fe/H] < 0.3$ (b). Red dashed line shows the [C/N] age calibration from \citet{Spoo_2022}}
    %     \label{fig:CNvAge}
    % \end{figure*}

    \subsection{Validation}
\begin{figure}[ht!]
        \centering
        \includegraphics[width=0.5\textwidth]{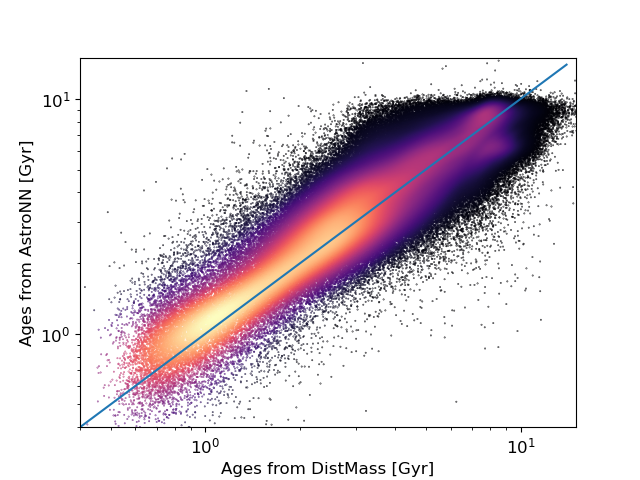}
        \caption{Distmass ages vs \protect\citet{2019MNRAS.489..176M} ages for stars within the bounded training region for the mass model}
        \label{fig:ageComparisonBounded}
    \end{figure}

    The AstroNN catalog contains stellar ages calculated by \citet{2019MNRAS.489..176M} which are derived using a Bayesian convolutional neural network trained on the full APOGEE spectra and asteroseismic ages, as opposed to our method which uses the derived stellar parameters and abundances. The comparison to \citet{2019MNRAS.489..176M} is not an independent validation of our ages since they use the same asteroseismic training set, but they do serve as a quick sanity check for our ages. Figure \ref{fig:ageComparisonBounded} compares our results with the AstroNN results, using only stars within the parameter space covered by the training set of the mass model shown in Figure \ref{fig:HRMass}.  The figure shows that our age predictions are generally in agreement with the ages from \citet{2019MNRAS.489..176M}. 

    % Star clusters are composed of stars that all should have largely the same age, so they can act as a tool to indicate prediction reliability. Figure \ref{fig:ageComparisonBoundedCluster} shows our predicted ages vs those from \citet{2019MNRAS.489..176M} for stars within clusters. While both figures \ref{fig:ageComparisonBounded} and \ref{fig:ageComparisonBoundedCluster} show that our age predictions generally agree with the age estimations from \citet{2019MNRAS.489..176M}, the scatter in age predictions is high and not all the clusters have tight age predictions. The average standard deviation of our star age predictions in the clusters is 0.16. For some clusters, such as N6791 and M67, our predictions show self consistent ages with standard deviations $<0.1$, but others, such as N6388 and N7789, have a spread of age predictions with standard deviations $>0.2$ or have a bimodal distribution of ages. 
    
\begin{figure}[ht!]
        \centering
        \includegraphics[width=0.8\textwidth]{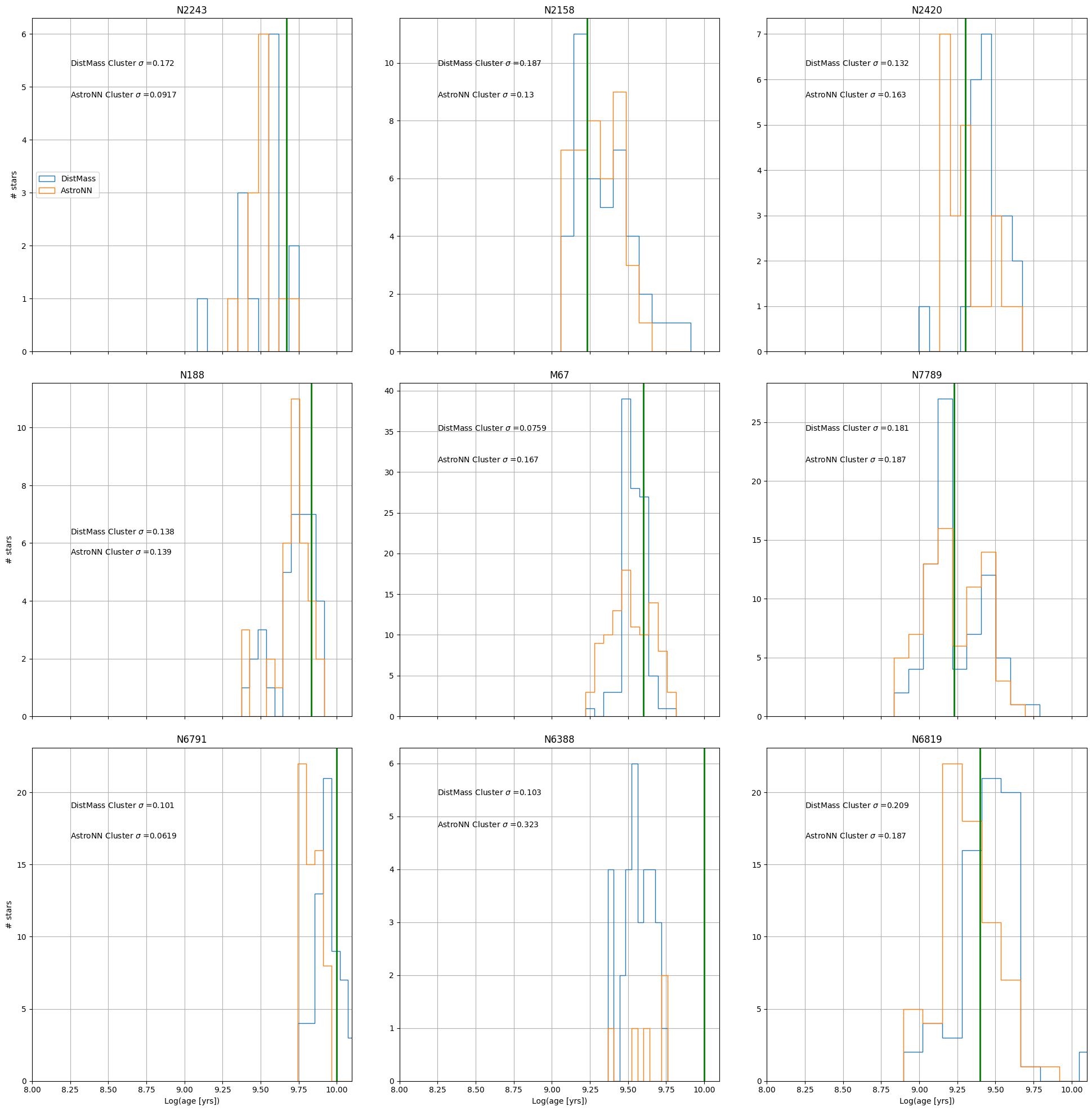}
        \caption{Distmass and \protect\citet{2019MNRAS.489..176M} age histograms for stars in clusters. The DistMass ages used here are from the corrected Mosser seismology}
        \label{fig:ageComparisonBoundedCluster}
    \end{figure}

    Star clusters are composed of stars that all should have largely the same age, so they can act as a tool to indicate prediction reliability. Figure \ref{fig:ageComparisonBoundedCluster} shows age predictions from our model and \citet{2019MNRAS.489..176M} for a selection of stellar clusters along with the scatter of age predictions for each cluster by the models. The star cluster comparisons in Figure \ref{fig:ageComparisonBoundedCluster} show that our model has good agreement with \citet{2019MNRAS.489..176M}, and have fairly tight age distributions within each cluster.

    Some of the clusters shown in Figure \ref{fig:ageComparisonBoundedCluster} have inconsistencies with the literature age. NGC 6388 is predicted by both our ages and \citet{2019MNRAS.489..176M} to be younger than the listed age we used for N6388, with an age of 3.65 Gyr. The literature age for NGC 6388 is from \citet{2006ApJ...651L.133C} who notes the age relative to the age of the cluster 47 TUC as $\Delta t \approx -0.5 \pm 1.6$ Gyr (NGC 6388 being younger). We used the age of 47 TUC from \citet{10.1093/mnras/stx378} with the information from \citet{2006ApJ...651L.133C} to calculate an age of 11.3 Gyr for NGC 6388. Given that N6388 is a globular cluster, the young age predicted by our model and \citet{2019MNRAS.489..176M} is likely to be an underestimate and the literature age is closer to reality. A plausible explanation for the discrepancy is there are multiple populations of stars in N6388. \citet{2020MNRAS.492.1641M} notes an N-C anti-correlation in many globular clusters including NGC 6388 which is likely due to pollution from an older population of FG stars. The result of the pollution is a lower birth [C/N] for a large population of stars which would appear as the mass/age models under-predicting the ages of the stars. 

\begin{figure}[ht!]
        \centering
        \includegraphics[width=0.8\textwidth]{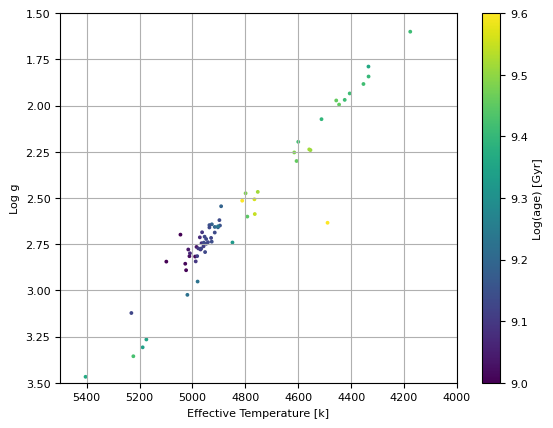}
        \caption{kiel diagram showing stars in NGC 7789 color coded by Distmass. The DistMass ages used here are from the corrected Mosser seismology}
        \label{fig:N7789}
    \end{figure}

    Another cluster we examined, N7789, has a bimodel distribution of predicted ages for both our method and \citet{2019MNRAS.489..176M}. Figure \ref{fig:N7789} shows a kiel diagram of stars in NGC 7789 color coded by Distmass age. We see a lower age prediction in the red clump, which means that the mass prediction is higher in that region. N7789 is the only cluster we examined that had this feature. Confirming with APOKASC asteroseismology is not an option as there are few stars in N7789 with APOKASC measurements, although additional asteroseismology from K2 and TESS may help address this issue. Red clump stars do experience mass-loss as described in section \ref{sec: mass}, however mass loss would result in an age over-prediction, while the predicted ages for red clump stars in N7789 are younger. \citet{2010IAUS..266..320G} suggests there are peculiarities in the red clump of N7789, specifically that there may be a faint secondary red clump.

\section{Conclusion}\label{conc}

{\scriptsize\begin{deluxetable}{ccccc}
\tablenum{2}
\tablecaption{DistMass catalog data contents\label{tab:datamodel}}
\tablewidth{0pt}
\tablehead{
\colhead{Data} & \colhead{header name}  &\nocolhead{Common}& \colhead{Description} & \colhead{source} 
}
\startdata
APOGEE ID & APOGEE\_ID & & ID from APOGEE  & SDSS DR17  \\
APSTAR ID & APSTAR\_ID && ID from APSTAR  & SDSS DR17  \\
Effective Temperature & TEFF && Stellar effective temperature  & SDSS DR17 \\
Surface Gravity & LOGG && Surface gravity in log space  & SDSS DR17  \\
Metallicity & M\_H  && Stellar metallicity  & SDSS DR17  \\
Carbon Abundance & C\_FE && Stellar carbon abundance  & SDSS DR17  \\
Nitrogen Abundance & N\_FE && Stellar nitrogen abundance  & SDSS DR17  \\
Gaia Distance & GAIAEDR3\_DIST & &Distance from Gaia used for training  & Gaia DR3 \\
Extinction & EXTINCTION && Extinction magnitude used as described in section \ref{sec:setcreation}  & SDSS DR17  \\
Apparent Mag & MAG && Apparent K magnitude from APOGEE   & SDSS DR17  \\
Absolute Mag & ABS\_MAG && The predicted absolute magnitude from our model  & DistMass  \\
Absolute Mag error & ABS\_MAG\_ERR && estimated error in our magnitude prediction   & DistMass  \\
DistMass Distance & DISTANCE && Predicted distance from our model  & DistMass  \\
DistMass Distance error & DISTANCE\_ERR && Error in our predicted distance   & DistMass  \\
Weighted DistMass Distance & DISTANCE\_WEIGHTED && DistMass distances weighted with Gaia distances  & DistMass  \\
DistMass mass uncorrected Sharma & MASS\_UNCOR\_SS && Estimated mass from the uncorrected Sharma model  & DistMass  \\
DistMass mass corrected Sharma & MASS\_COR\_SS && Corrected Sharma model mass & DistMass  \\
DistMass mass uncorrected Mosser & MASS\_UNCOR\_MO && Uncorrected Mosser model mass & DistMass  \\
DistMass mass corrected Mosser & MASS\_COR\_MO && Corrected Mosser model mass & DistMass  \\
DistMass mass uncorrected White & MASS\_UNCOR\_TW && Uncorrected White model mass & DistMass  \\
DistMass mass corrected White & MASS\_COR\_TW && Corrected White model mass & DistMass  \\
DistMass Mass error & MASS\_ERR && Mass error estimate for all models   & DistMass  \\
DistMass age uncorrected Sharma & AGE\_UNCOR\_SS && Estimated age from our uncorrected Sharma model  & DistMass  \\
DistMass age corrected Sharma & AGE\_COR\_SS && Estimated age from the corrected Sharma model  & DistMass  \\
DistMass age uncorrected Mosser & AGE\_UNCOR\_MO && Estimated age from the uncorrected Mosser model  & DistMass  \\
DistMass age corrected Mosser & AGE\_COR\_MO && Estimated age from the corrected Mosser model  & DistMass  \\
DistMass age uncorrected White & AGE\_UNCOR\_TW && Estimated age from the uncorrected White model  & DistMass  \\
DistMass age corrected White & AGE\_COR\_TW && Estimated age from the corrected White model  & DistMass  \\
DistMass age error  &  AGE\_ERR && Estimated error in our age estimates for all models  & DistMass \\
BITMASK & BITMASK && Contains flags to indicate notes about given star &    \\
NMSU distance & NMSU\_DIST && Distances from method described in \ref{nmsudist} & DistMass  \\
\enddata
\end{deluxetable}}

\begin{deluxetable*}{cchlDlc}
\tablenum{3}
\tablecaption{Bitmask description\label{tab:bitmask}}
\tablewidth{0pt}
\tablehead{
\colhead{Bit} & \colhead{Bit location} &\nocolhead{Common} &  \colhead{Description}
}
\startdata
1 & 000001 && Marks stars that were members of distance model training set \\
2 & 000010 && Stars outside the covered parameter space of the mass model \\
3 & 001000 && Training mass from APOKASC \citep{2018} \\
4 & 010000 && Training mass from \citet{2020} \\
5 & 100000 && Training mass from \citet{2013MNRAS.429.3645S} \\
\enddata
\label{tab:bitmask}
\end{deluxetable*}

Simple neural networks were used to derive distances and masses for stars in SDSS Data Release 17. We also used the predicted masses to derive ages for select giant stars. Our distance model derived distances for 733901 unique spectra with good coverage of parameter space with the exception of very low log g and metal poor stars. These distance predictions have an uncertainty of 20\% for stars at log g $< 2$as determined from comparison with the Gaia training set, and uncertainties below 10\% for stars with log g $> 2.5$. Comparisons of distance predictions to clusters indicates that the model has a median fractional error of -6\% and a scatter of 11.4\%.

The mass model was also applied to all the stars in the dataset, however, the training parameter space is sparser, with a limited number of stars at log g $< 1.5$ and $[Fe/H] < -0.5$, which limits the set of stars the mass model provides valid predictions for. We derived masses for 506817 unique spectra that are inside the limited coverage of the training parameter space and we find that the mass model has median fractional errors of $<0.12$.

Stellar age prediction was performed using the mass model predictions and stellar isochrones. Our age prediction is only possible for giant stars since it relies on the assumption that a giant star is in the final 1 - 10\% of its lifespan. Since it also relies on the mass model predictions, it is further limited to stars within the mass model training parameter space. In total, we have age predictions for 271615 unique spectra. To examine the validity of the age predictions we compared the ages of stars inside clusters and found that the average standard deviation of age in log(age) is $<0.1$. Some clusters had much higher scatter, or in the case of N7789, an unexplained bimodel distribution of ages. However, most clusters we examined showed self-consistent age predictions or had plausible known reasons for having high scatter. We also presented some initial examinations of the C/N - age relation using our data compared to work from \citet{Spoo_2022}. 

All of the derived distances, masses, and ages are available in a Value Added Catalog for SDSS Data Release 17
\footnote{\href {https://www.sdss4.org/dr17/data_access/value-added-catalogs/?vac_id=distmass:-distances,-masses,-and-ages-for-apogee-dr17}{www.sdss4.org/dr17/data\_access/value-added-catalogs/?vac\_id=distmass:-distances,-masses,-and-ages-for-apogee-dr17}} \citep{2022ApJS..259...35A}. The catalog provides all the input parameters and training labels used by the models, such as effective temperature, surface gravity, and metallicity. Table \ref{tab:datamodel} shows the complete contents of the catalog. A bitmask is provided which contains flags for indicating whether a star was a member of the training set, is outside the training parameter space for the mass model, and the source of the mass label; see Table \ref{tab:bitmask}.

Potential uses of the catalog could include testing stellar mixing models, examining various star clusters, probing the history of the Milky Way, etc. \citet{Imig_2023} use the distances and ages from our catalog to produce chemical and age maps of the Milky Way. Future work will include using our methods to produce a catalog utilizing new asteroseismic masses from K2 and TESS for the SDSS-V Milky Way Mapper project.

\section{ACKNOWLEDGEMENTS}

A.S-M., J.A.H., and J.I gratefully acknowledge support from NSF grant AST-1909897.

Funding for the Sloan Digital Sky 
Survey IV has been provided by the 
Alfred P. Sloan Foundation, the U.S. 
Department of Energy Office of 
Science, and the Participating 
Institutions. 

SDSS-IV acknowledges support and 
resources from the Center for High 
Performance Computing  at the 
University of Utah. The SDSS 
website is www.sdss.org.

SDSS-IV is managed by the 
Astrophysical Research Consortium 
for the Participating Institutions 
of the SDSS Collaboration including 
the Brazilian Participation Group, 
the Carnegie Institution for Science, 
Carnegie Mellon University, Center for 
Astrophysics | Harvard \& 
Smithsonian, the Chilean Participation 
Group, the French Participation Group, 
Instituto de Astrof\'isica de 
Canarias, The Johns Hopkins 
University, Kavli Institute for the 
Physics and Mathematics of the 
Universe (IPMU) / University of 
Tokyo, the Korean Participation Group, 
Lawrence Berkeley National Laboratory, 
Leibniz Institut f\"ur Astrophysik 
Potsdam (AIP),  Max-Planck-Institut 
f\"ur Astronomie (MPIA Heidelberg), 
Max-Planck-Institut f\"ur 
Astrophysik (MPA Garching), 
Max-Planck-Institut f\"ur 
Extraterrestrische Physik (MPE), 
National Astronomical Observatories of 
China, New Mexico State University, 
New York University, University of 
Notre Dame, Observat\'ario 
Nacional / MCTI, The Ohio State 
University, Pennsylvania State 
University, Shanghai 
Astronomical Observatory, United 
Kingdom Participation Group, 
Universidad Nacional Aut\'onoma 
de M\'exico, University of Arizona, 
University of Colorado Boulder, 
University of Oxford, University of 
Portsmouth, University of Utah, 
University of Virginia, University 
of Washington, University of 
Wisconsin, Vanderbilt University, 
and Yale University.

% \appendix
% \section{Appendix information}

% *I'm thinking of placing all the cluster plots I made here*

\bibliography{references}{}
\bibliographystyle{aasjournal}

\end{document}